\documentclass[aps,prd,twocolumn,superscriptaddress,showpacs,preprintnumbers,amsmath,amssymb,floatfix]{revtex4-1}

\usepackage[dvipdfmx]{graphicx}
\usepackage{graphicx}
\usepackage{dcolumn}
\usepackage{color}
\usepackage[mathlines]{lineno}
\usepackage{tabularx}
\usepackage{tikz}       
\usepackage{tikz-feynman}
\usepackage{mathtools}
\usepackage{multirow}
\usepackage[titletoc,toc,title]{appendix}
\graphicspath{{ps}}
\usepackage{float}

\definecolor{BlueViolet}{rgb}{0.2, 0.00, 0.7}
\definecolor{Blue}{rgb}{0.15, 0.00, 0.9}
\usepackage[colorlinks=true, linkcolor=Blue, citecolor=Blue, urlcolor=BlueViolet]{hyperref} 
\renewcommand{\arraystretch}{1.1}
\newcolumntype{Y}{>{\centering\arraybackslash}X}

\begin{document}
\preprint{\vbox{ \hbox{   }
    \hbox{Belle Preprint {2021-25}}
    \hbox{KEK Preprint {2021-30}}
}}
\title{Study of \boldmath{$\overline{B}{}^0\rightarrow D^{+}h^{-} (h=K/\pi)$} decays at Belle }
\noaffiliation
\affiliation{Department of Physics, University of the Basque Country UPV/EHU, 48080 Bilbao}
\affiliation{University of Bonn, 53115 Bonn}
\affiliation{Brookhaven National Laboratory, Upton, New York 11973}
\affiliation{Budker Institute of Nuclear Physics SB RAS, Novosibirsk 630090}
\affiliation{Faculty of Mathematics and Physics, Charles University, 121 16 Prague}
\affiliation{Chonnam National University, Gwangju 61186}
\affiliation{University of Cincinnati, Cincinnati, Ohio 45221}
\affiliation{Deutsches Elektronen--Synchrotron, 22607 Hamburg}
\affiliation{University of Florida, Gainesville, Florida 32611}
\affiliation{Department of Physics, Fu Jen Catholic University, Taipei 24205}
\affiliation{Key Laboratory of Nuclear Physics and Ion-beam Application (MOE) and Institute of Modern Physics, Fudan University, Shanghai 200443}
\affiliation{Justus-Liebig-Universit\"at Gie\ss{}en, 35392 Gie\ss{}en}
\affiliation{Gifu University, Gifu 501-1193}
\affiliation{SOKENDAI (The Graduate University for Advanced Studies), Hayama 240-0193}
\affiliation{Gyeongsang National University, Jinju 52828}
\affiliation{Department of Physics and Institute of Natural Sciences, Hanyang University, Seoul 04763}
\affiliation{University of Hawaii, Honolulu, Hawaii 96822}
\affiliation{High Energy Accelerator Research Organization (KEK), Tsukuba 305-0801}
\affiliation{J-PARC Branch, KEK Theory Center, High Energy Accelerator Research Organization (KEK), Tsukuba 305-0801}
\affiliation{National Research University Higher School of Economics, Moscow 101000}
\affiliation{Forschungszentrum J\"{u}lich, 52425 J\"{u}lich}
\affiliation{IKERBASQUE, Basque Foundation for Science, 48013 Bilbao}
\affiliation{Indian Institute of Science Education and Research Mohali, SAS Nagar, 140306}
\affiliation{Indian Institute of Technology Guwahati, Assam 781039}
\affiliation{Indian Institute of Technology Hyderabad, Telangana 502285}
\affiliation{Indian Institute of Technology Madras, Chennai 600036}
\affiliation{Indiana University, Bloomington, Indiana 47408}
\affiliation{Institute of High Energy Physics, Chinese Academy of Sciences, Beijing 100049}
\affiliation{Institute of High Energy Physics, Vienna 1050}
\affiliation{Institute for High Energy Physics, Protvino 142281}
\affiliation{INFN - Sezione di Napoli, I-80126 Napoli}
\affiliation{INFN - Sezione di Roma Tre, I-00146 Roma}
\affiliation{INFN - Sezione di Torino, I-10125 Torino}
\affiliation{Advanced Science Research Center, Japan Atomic Energy Agency, Naka 319-1195}
\affiliation{J. Stefan Institute, 1000 Ljubljana}
\affiliation{Institut f\"ur Experimentelle Teilchenphysik, Karlsruher Institut f\"ur Technologie, 76131 Karlsruhe}
\affiliation{Kavli Institute for the Physics and Mathematics of the Universe (WPI), University of Tokyo, Kashiwa 277-8583}
\affiliation{Department of Physics, Faculty of Science, King Abdulaziz University, Jeddah 21589}
\affiliation{Kitasato University, Sagamihara 252-0373}
\affiliation{Korea Institute of Science and Technology Information, Daejeon 34141}
\affiliation{Korea University, Seoul 02841}
\affiliation{Kyoto Sangyo University, Kyoto 603-8555}
\affiliation{Kyungpook National University, Daegu 41566}
\affiliation{Universit\'{e} Paris-Saclay, CNRS/IN2P3, IJCLab, 91405 Orsay}
\affiliation{P.N. Lebedev Physical Institute of the Russian Academy of Sciences, Moscow 119991}
\affiliation{Faculty of Mathematics and Physics, University of Ljubljana, 1000 Ljubljana}
\affiliation{Ludwig Maximilians University, 80539 Munich}
\affiliation{Luther College, Decorah, Iowa 52101}
\affiliation{Malaviya National Institute of Technology Jaipur, Jaipur 302017}
\affiliation{Faculty of Chemistry and Chemical Engineering, University of Maribor, 2000 Maribor}
\affiliation{Max-Planck-Institut f\"ur Physik, 80805 M\"unchen}
\affiliation{School of Physics, University of Melbourne, Victoria 3010}
\affiliation{University of Mississippi, University, Mississippi 38677}
\affiliation{University of Miyazaki, Miyazaki 889-2192}
\affiliation{Moscow Physical Engineering Institute, Moscow 115409}
\affiliation{Graduate School of Science, Nagoya University, Nagoya 464-8602}
\affiliation{Kobayashi-Maskawa Institute, Nagoya University, Nagoya 464-8602}
\affiliation{Universit\`{a} di Napoli Federico II, I-80126 Napoli}
\affiliation{Nara Women's University, Nara 630-8506}
\affiliation{National Central University, Chung-li 32054}
\affiliation{National United University, Miao Li 36003}
\affiliation{Department of Physics, National Taiwan University, Taipei 10617}
\affiliation{H. Niewodniczanski Institute of Nuclear Physics, Krakow 31-342}
\affiliation{Nippon Dental University, Niigata 951-8580}
\affiliation{Niigata University, Niigata 950-2181}
\affiliation{Novosibirsk State University, Novosibirsk 630090}
\affiliation{Osaka City University, Osaka 558-8585}
\affiliation{Pacific Northwest National Laboratory, Richland, Washington 99352}
\affiliation{Panjab University, Chandigarh 160014}
\affiliation{Peking University, Beijing 100871}
\affiliation{University of Pittsburgh, Pittsburgh, Pennsylvania 15260}
\affiliation{Punjab Agricultural University, Ludhiana 141004}
\affiliation{Research Center for Nuclear Physics, Osaka University, Osaka 567-0047}
\affiliation{Meson Science Laboratory, Cluster for Pioneering Research, RIKEN, Saitama 351-0198}
\affiliation{Dipartimento di Matematica e Fisica, Universit\`{a} di Roma Tre, I-00146 Roma}
\affiliation{Department of Modern Physics and State Key Laboratory of Particle Detection and Electronics, University of Science and Technology of China, Hefei 230026}
\affiliation{Showa Pharmaceutical University, Tokyo 194-8543}
\affiliation{Soongsil University, Seoul 06978}
\affiliation{Sungkyunkwan University, Suwon 16419}
\affiliation{School of Physics, University of Sydney, New South Wales 2006}
\affiliation{Department of Physics, Faculty of Science, University of Tabuk, Tabuk 71451}
\affiliation{Tata Institute of Fundamental Research, Mumbai 400005}
\affiliation{Department of Physics, Technische Universit\"at M\"unchen, 85748 Garching}
\affiliation{School of Physics and Astronomy, Tel Aviv University, Tel Aviv 69978}
\affiliation{Toho University, Funabashi 274-8510}
\affiliation{Department of Physics, Tohoku University, Sendai 980-8578}
\affiliation{Earthquake Research Institute, University of Tokyo, Tokyo 113-0032}
\affiliation{Department of Physics, University of Tokyo, Tokyo 113-0033}
\affiliation{Tokyo Institute of Technology, Tokyo 152-8550}
\affiliation{Tokyo Metropolitan University, Tokyo 192-0397}
\affiliation{Virginia Polytechnic Institute and State University, Blacksburg, Virginia 24061}
\affiliation{Wayne State University, Detroit, Michigan 48202}
\affiliation{Yamagata University, Yamagata 990-8560}
\affiliation{Yonsei University, Seoul 03722}
  \author{E.~Waheed}\affiliation{High Energy Accelerator Research Organization (KEK), Tsukuba 305-0801} 
 \author{P.~Urquijo}\affiliation{School of Physics, University of Melbourne, Victoria 3010} 
  \author{I.~Adachi}\affiliation{High Energy Accelerator Research Organization (KEK), Tsukuba 305-0801}\affiliation{SOKENDAI (The Graduate University for Advanced Studies), Hayama 240-0193} 
  \author{H.~Aihara}\affiliation{Department of Physics, University of Tokyo, Tokyo 113-0033} 
  \author{S.~Al~Said}\affiliation{Department of Physics, Faculty of Science, University of Tabuk, Tabuk 71451}\affiliation{Department of Physics, Faculty of Science, King Abdulaziz University, Jeddah 21589} 
  \author{D.~M.~Asner}\affiliation{Brookhaven National Laboratory, Upton, New York 11973} 
  \author{H.~Atmacan}\affiliation{University of Cincinnati, Cincinnati, Ohio 45221} 
  \author{V.~Aulchenko}\affiliation{Budker Institute of Nuclear Physics SB RAS, Novosibirsk 630090}\affiliation{Novosibirsk State University, Novosibirsk 630090} 
  \author{T.~Aushev}\affiliation{National Research University Higher School of Economics, Moscow 101000} 
 \author{S.~Bahinipati}\affiliation{Indian Institute of Technology Bhubaneswar, Satya Nagar 751007} 
% ...
\collaboration{The Belle Collaboration}
\begin{abstract}
We present a measurement of the branching fractions of the Cabibbo favored $\overline{B}{}^0\rightarrow D^{+}\pi^{-}$ and  the Cabibbo suppressed $\overline{B}{}^0\rightarrow D^{+}K^{-}$  decays. We find $\mathcal{B}(\overline{B}{}^0\rightarrow D^{+}\pi^{-}) = (2.48 \pm 0.01 \pm 0.09 \pm 0.04) \times 10^{-3}$ and $\mathcal{B}(\overline{B}{}^0\rightarrow D^{+}K^{-})= (2.03 \pm 0.05 \pm 0.07 \pm 0.03) \times 10^{-4}$ decays, where the first uncertainty is statistical,
the second is systematic, and the third uncertainty is due to the $D^{+} \rightarrow K^{-}\pi^{+}\pi^{+}$ branching fraction. The ratio of branching fractions of $\overline{B}{}^0\rightarrow D^{+}K^{-}$ and $\overline{B}{}^0\rightarrow D^{+}\pi^{-}$ is measured to be $R^{D} = [8.19 \pm 0.20(\rm stat) \pm 0.23(\rm  syst)] \times 10^{-2} $. These measurements are performed using the full Belle dataset, which corresponds to $772 \times 10^{6} B\overline{B} $ pairs and use the Belle II software framework for data analysis.
\end{abstract}
\maketitle
\section{Introduction}
\label{sec:Intro}
Two-body decays of $B$ mesons serve as an important test bed for phenomenological studies of the quark flavor sector of the Standard Model  of particle physics. The Cabibbo-favored mode $\overline{B}{}^0\rightarrow D^{+}\pi^{-}$  is an especially clean and  abundant hadronic decay that provides a good opportunity to test models of hadronic $B$ meson decays. Due to the large mass of the $b$ quark, the influence of  the strong interaction in these decays can be calculated more reliably than those in light-meson decays. It has been suggested that improved measurements of color-favored hadronic two-body decays of  $B$ mesons will lead to a better understanding of poorly known quantum chromodynamics (QCD) effects \cite{cite-qcdeffetcs}. The decays of $B$ mesons to two-body hadronic final states can be analyzed by decomposing their amplitudes in terms of different decay topologies and then applying SU(3) flavor symmetry of QCD to derive relations between them. The Cabibbo-suppressed mode $\overline{B}{}^0\rightarrow D^{+}K^{-}$ only receives contributions from color-allowed tree amplitudes while $\overline{B}{}^0\rightarrow D^{+}\pi^{-}$ receives contributions from both color-allowed tree and exchange amplitudes \cite{cite-arXiv:1012.2784}. These two decay modes can be related by a ratio \cite{cite-belleold},
\begin{eqnarray}
R^{D} &\equiv & \frac{\mathcal{B}(\overline{B}{}^0\rightarrow D^{+}K^{-})}{\mathcal{B}(\overline{B}{}^0\rightarrow D^{+}\pi^{-})}\simeq \tan^{2} \theta_{\rm C} \left(\frac{f_{K}}{f_{\pi}}\right)^2,
\end{eqnarray}
where $\theta_{\rm C}$ is the Cabibbo angle,  and $f_K$ and $f_\pi$ are meson decay constants. The theoretical description for these hadronic decays has considerably improved over the years \cite{cite-theory1, cite-theory2} and has been followed by several recent developments \cite{cite-arXix1606.02888, cite-arXiv:2007.10338}. This description relies on factorization and SU(3)-symmetry assumptions, so measurements of these modes can be used to test these hypotheses in heavy-quark hadronic decays. The above two modes are also important because they constitute high-statistics control samples for the hadronic $B$-decay measurements related to time-dependent $CP$ violation and the extraction of the Cabibbo-Kobayashi-Maskawa unitarity-triangle angle $\phi_{3}$ \cite{cite-theory3}. Experimentally, calculating the ratio of the branching fractions of $\overline{B}{}^0\rightarrow D^{+}K^{-}$ and $\overline{B}{}^0\rightarrow D^{+}\pi^{-}$ modes has the advantage that many systematic uncertainties cancel, enabling tests of theoretical predictions, particularly those of factorization and SU(3) symmetry breaking in QCD.

The theoretical predictions made in Refs. \cite{cite-arXix1606.02888, cite-arXiv:2007.10338}  are based on the framework of QCD factorization, at next-to-next-to-leading order. However, these predictions significantly differ from the experimental values. Several attempts \cite{cite-arXiv:2103.04138, cite-arXiv:2109.04950, cite-arXiv:2008.01086} have been made to explain the discrepancy in both $\overline{B}{}^0\rightarrow D^{+}\pi^{-}$ and $\overline{B}{}^0\rightarrow D^{+}K^{-}$ decays within the context of new physics. Final-state rescattering effects on  $\overline{B}{}^0\rightarrow D^{+}h^{-} (h={K/\pi})$ have also been proposed to explain the discrepancy \cite{cite-arXiv:2109.10811}. The results in Ref. \cite{cite-arXiv:2109.10811} rule out rescattering effects as a cause for the discrepancies and hence hint at a possible beyond-the-SM explanation.

Earlier, Belle reported a study of the Cabibbo-suppressed  $\overline{B}{}^0\rightarrow D^{+}K^{-}$ decay using a  small data datasetset \cite{cite-belleold} by measuring the ratio of branching fraction of Cabibbo-suppressed $\overline{B}{}^0\rightarrow D^{+}K^{-}$ to that of the Cabibbo-favored $\overline{B}{}^0\rightarrow D^{+}\pi^{-}$ decay. The branching fraction for $\overline{B}{}^0\rightarrow D^{+}\pi^{-}$ decay was previously measured by BABAR \cite{cite-BaBar1, cite-BaBar2}, CLEO \cite{cite-cleo1, cite-cleo2} and ARGUS \cite{cite-argus}. LHCb measured the branching fraction of $\overline{B}{}^0\rightarrow D^{+}K^{-}$ as well as the ratio of hadronization fractions $f_{s}/f_{d}$ \cite{cite-DKlhcb}. A clear understanding of $\overline{B}{}^0\rightarrow D^{+}h^{-} (h={K/\pi})$ decays constitutes an important ingredient for the measurement $f_{s}/f_{d}$, which in turn will aid  the measurement of rare decay $B_{s}^0\rightarrow \mu^{+}\mu^{-}$. Currently, the world averages \cite{cite-PDG} for the branching fractions of $\overline{B}{}^0\rightarrow D^{+}K^{-}$ and $\overline{B}{}^0\rightarrow D^{+}\pi^{-}$ decays are $\mathcal{B}(\overline{B}{}^0\rightarrow D^{+}K^{-}) = (1.86 \pm 0.20) \times 10^{-4}$ and $\mathcal{B}(\overline{B}{}^0\rightarrow D^{+}\pi^{-}) = (2.52 \pm 0.13) \times 10^{-3}$, respectively, where the uncertainty is the sum in quadrature of the statistical and systematic errors. LHCb \cite{cite-ratiolhcb} measured the ratio of the branching fractions for $\overline{B}{}^0\rightarrow D^{+}K^{-}$ and $\overline{B}{}^0\rightarrow D^{+}\pi^{-}$ to be $0.0822 \pm 0.0011(\rm stat) \pm 0.0025(\rm syst)$, which dominates the current world-average value.
  
In this paper, we present measurements of the branching fractions of $\overline{B}{}^0\rightarrow D^{+}\pi^{-}$ and $\overline{B}{}^0\rightarrow D^{+}K^{-}$ decays using the full $\Upsilon(4S)$ dataset collected with the Belle detector. 

The paper is organized as follows. Sec.~\ref{sec:detector} describes the Belle detector, as well as the data and simulation samples used in this analysis. The event selection requirements are outlined in Sec.~\ref{sec:reco}. Sec.~\ref{sec:simfit} describes how the values of $R^{D}$ and the $\overline{B}{}^0\rightarrow D^{+}h^{-}(h=K/\pi)$  branching fraction are determined from the data. The results and the evaluation of systematic uncertainties are described in Sec.~\ref{sec:results}, and the conclusion is given in Sec.~\ref{sec:conclusion}. 

\section{THE BELLE DETECTOR AND DATA SAMPLE}
\label{sec:detector}
We use the full $\Upsilon(4S)$ data sample containing $772 \times 10^6~B\overline{B}$  events recorded with the Belle detector~\cite{cite-Belle} at the KEKB asymmetric-beam-energy $e^+ e^-$ collider ~\cite{cite-KEKB}. Belle is a large-solid-angle magnetic spectrometer that consists of a silicon vertex detector, a 50-layer central drift chamber (CDC), an array of aerogel threshold Cherenkov counters (ACC), a barrel-like arrangement of time-of-flight scintillation counters (TOF), and an electromagnetic calorimeter comprised of CsI(Tl) crystals. All these detector components are located inside a superconducting solenoid coil that provides a 1.5~T magnetic field~\cite{cite-Belle}. 

A Monte Carlo (MC) simulated event sample is used to optimize the event selection, study background and compare the distributions observed in collision data with expectations. A signal-only simulated event sample is utilized to model the features of the signal for fits and determine selection efficiencies. One million signal events are generated for both decay channels.  The so-called generic MC sample consists of simulated events that include $e^+ e^- \to$ $B\overline{B}{}$, $u\overline{u}$, $d\overline{d}$, $s\overline{s}$, and $c\overline{c}$  processes in realistic proportions, and corresponds in size to more than five times the $\Upsilon(4S)$ data. The generic MC sample is used to study background and make comparisons with the data. The $B$- and $D$-meson decays are simulated with the \textsc{EvtGen} generator \cite{cite-EvtGen} where the D\_DALITZ model is used for the $D^{+}\rightarrow K^{-}\pi^{+} \pi^{+}$ final state. The effect of final-state radiation is simulated by the \textsc{PHOTOS} package \cite{cite-PHOTOS}. The interactions of particles with the detector are simulated using \textsc{GEANT3} \cite{cite-GEANT}.

\section{EVENT SELECTION AND RECONSTRUCTION}
\label{sec:reco}
We use the Belle II Analysis Software Framework (basf2) \cite{cite-basf2} for the decay-chain reconstruction and convert the Belle data to basf2 format using the B2BII software package \cite{cite-b2bii}. The decays $\overline{B}{}^0\rightarrow D^{+}\pi^{-}$ and $\overline{B}{}^0\rightarrow D^{+}K^{-}$ have nearly the same kinematic properties. The former is used to establish selection criteria on kinematic variables and  determine the experimental resolution due to its larger data size compared to the latter. Charged particle tracks originating from $e^{+}e^{-}$ collisions are selected by requiring $dr<0.2~\rm cm $ and $|dz|<1.5~\rm cm$, where $dr$ and $|dz|$ represent the distance of closest approach to the interaction point (IP) in the plane transverse to and along the $z$ axis, respectively. The $z$ axis is the direction opposite the $e^{+}$ beam.

\begin{figure}[ht!]
\centering
 \includegraphics[scale=0.45]{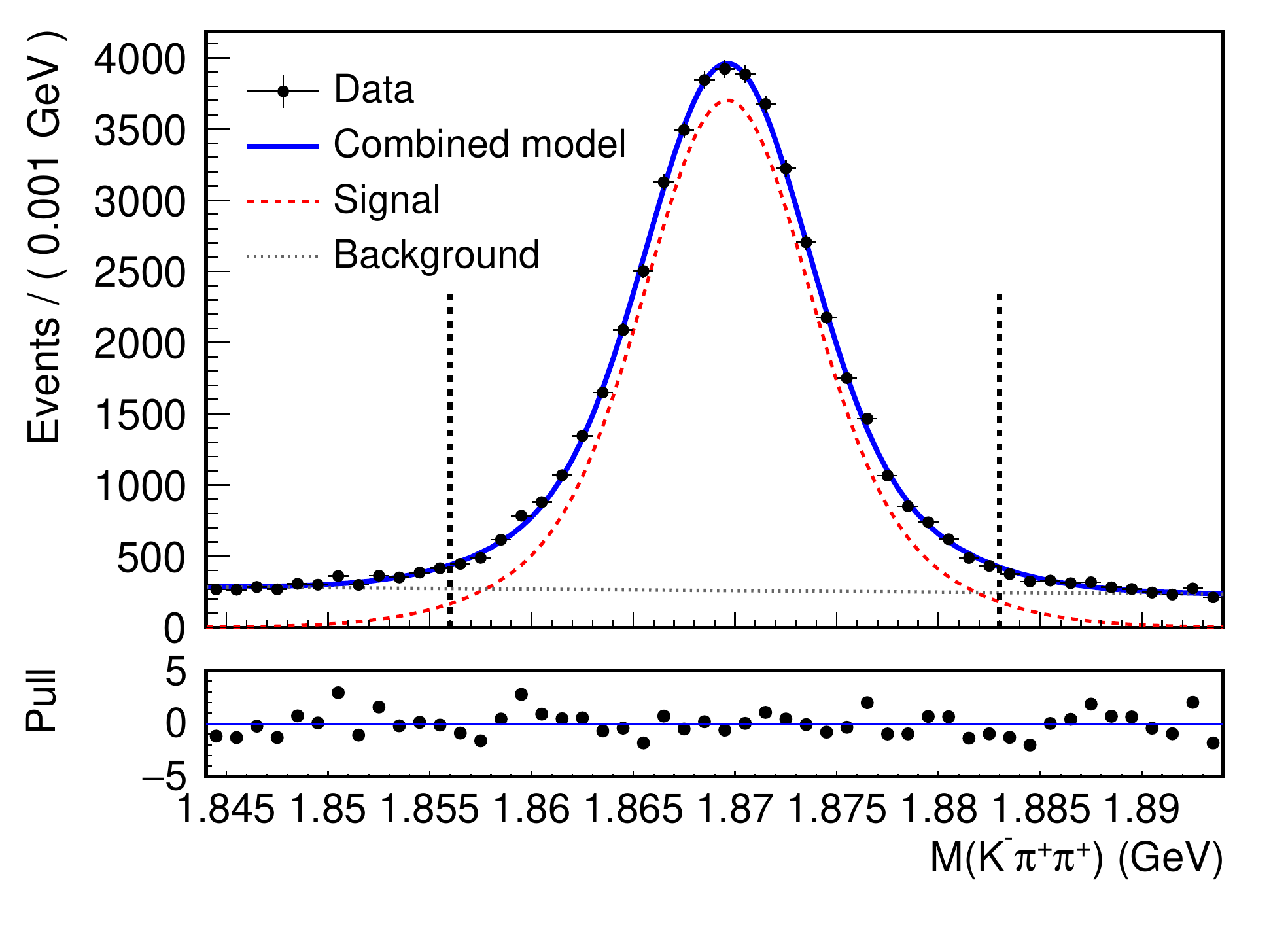} 
 \caption{Fit to the invariant mass distribution for $D^{+}\rightarrow K^{-}\pi^{+} \pi^{+}$ in data. The black vertical dotted lines show the $D$ mass window. The dashed curve shows the signal  component and dotted black line shows the background component. The distribution of pulls between the fit and the data points is also shown.}
 \label{fig:massD}
 \end{figure}

Information from the CDC, ACC, and TOF is used to determine a $K/\pi$ likelihood ratio $\mathcal{L}(K/\pi)= \frac{\mathcal{L}_{K}}{\mathcal{L}_{K}+\mathcal{L}_{\pi}}$ for charged particle identification (PID), where $\mathcal{L}_{K}$ and $\mathcal{L}_{\pi}$ are the likelihoods that a particular track is either a kaon or a pion, respectively. The likelihood value ranges from 0 to 1 where 0 (1) means the track is likely to be a $\pi$ ($K$).  To ensure high efficiency and  purity, we require  $\mathcal{L}(K/\pi)>0.6$ for kaon candidates and $\mathcal{L}(K/\pi)<0.6$  for pion candidates. The charged $D^{+}$ candidate is formed using $K^{-} \pi^{+} \pi^{+}$ combinations, which is then combined with a prompt hadron $(h = K/\pi)$ to form a $\overline{B}{}^0$ candidate. (The inclusion of charge conjugate states is implied throughout this paper.) $D^{+}$ meson candidates are  required to have a mass within $\pm 2.5\sigma$ of the known $D^{+}$ mass value \cite{cite-PDG}, where the Gaussian resolution $\sigma$ is approximately $5~ \rm MeV$. The effective $\sigma$ value is obtained by fitting the invariant mass distribution of $D^{+}\rightarrow K^{-}\pi^{+} \pi^{+}$ decays with a double Gaussian function for signal and a first-order polynomial for background as shown in Fig.~\ref{fig:massD}. 

The kinematic variables used to discriminate $B$ decays from background are the beam-energy-constrained mass 
\begin{eqnarray}
M_{\rm bc} &\equiv& \sqrt{E^{2}_{\rm beam}-p^{2}_{B}}, 
\end{eqnarray}
and the energy difference
\begin{eqnarray}
\Delta E &\equiv & E_{B} - E_{\rm beam}.
\end{eqnarray}
Here  $E_{B}$ and $p_{B}$ are the $B$ candidate's  energy and momentum, respectively,  and $E_{\rm beam}$ is the beam energy; these quantities are calculated in the $e^{+} e^{-}$ center-of-mass frame. Natural units $\hbar$ = c = 1 are used throughout the paper. For correctly reconstructed signal events, $M_{\rm bc}$ peaks at the known mass of the $\overline{B}{}^0$ meson  and $\Delta E$ peaks at zero. We retain the $\overline{B}{}^0$ candidates  satisfying $M_{\rm bc}>~5.27~\rm GeV$ and $|\Delta E|<0.13~\text{GeV}$.

The background from $e^{+}e^{-} \rightarrow q\overline{q}$ $(q = u,d,s,c)$ continuum processes are suppressed by requiring the ratio of the second-to-zeroth order Fox-Wolfram moments \cite{cite-r2} to be less than 0.3. This selection removes $\sim$$70\%$ of the continuum while rejecting $\sim$$30\%$ of the signal in both $\overline{B}{}^0\rightarrow D^{+}\pi^{-}$ and $\overline{B}{}^0\rightarrow D^{+}K^{-}$ decays. After applying the aforementioned selection criteria, only $0.7\%$ of events are found to have more than one candidate. In such events, we choose the best candidate as the one having the smallest value of $|M_{\rm bc} - m_B|$ where $m_B$ is the known $\overline{B}{}^0$ mass.
The kaon identification efficiency $\epsilon_{K}$ is determined from a kinematically selected sample of high momentum $D^{*+}$ mesons, which is used to calibrate the PID performance. With the application of the requirements $\mathcal{L}(K/\pi)<0.6$ for pions and $\mathcal{L}(K/\pi)>0.6$ for kaons, the kaon efficiency ($\epsilon_{K}$) value is found to  be $ (84.48 \pm 0.35)\%$ and the rate of pions misidentified as kaons is $(7.62 \pm 0.44)\%$.

 \section{Simultaneous Fit} 
 \label{sec:simfit}
As the $\overline{B}{}^0\rightarrow D^{+}\pi^{-}$ branching fraction is an order of magnitude larger than that of $\overline{B}{}^0\rightarrow D^{+}K^{-}$, the former can serve as an excellent calibration sample for the signal determination procedure. Furthermore, there is a significant contamination from $\overline{B}{}^0\rightarrow D^{+}\pi^{-}$ decays in the $\overline{B}{}^0\rightarrow D^{+}K^{-}$ sample in which the fast charged pion is misidentified as a kaon. A simultaneous fit to samples enriched in prompt tracks that are identified as either pions $[\mathcal{L}(K/\pi)<0.6]$ or kaons [$\mathcal{L}(K/\pi)>0.6$], allows us to directly determine this cross feed contribution from data. An unbinned maximum-likelihood fit is performed to extract the signal yield by fitting the $\Delta E$ distribution simultaneously in pion and kaon enriched samples. The yields of the $\overline{B}{}^0\rightarrow D^{+}\pi^{-}$ and $\overline{B}{}^0\rightarrow D^{+}K^{-}$ signals, as well as their cross feed contributions, in the pion and kaon enriched samples can be expressed by the following relations:
\begin{eqnarray}
N^{D^{+}\pi^{-}}_{\text{pion enhanced}} &=& (1-\kappa) \, N^{D^{+}\pi^{-}}_{\rm total},\\ 
N^{D^{+}\pi^{-}}_{\text{kaon enhanced}} &=& \kappa \, N^{D^{+}\pi^{-}}_{\rm total}, \\
N^{D^{+}K^{-}}_{\text{kaon enhanced}} &=& \epsilon_{K} \,R^{D}\, N^{D^{+}\pi^{-}}_{\rm total},\\ 
N^{D^{+}K^{-}}_{\text{pion enhanced}} &=& (1-\epsilon_{K})\,R^{D} \,N^{D^{+}\pi^{-}}_{\rm total}.
\end{eqnarray}
Here the values of $N^{D^{+}h^{-}}_{\text{pion enhanced}}(h=K/\pi)$ are the kaon and the pion yields in pion enriched sample with [$\mathcal{L}(K/\pi)<0.6$], and the $N^{D^{+}h^{-}}_{\text{kaon enhanced}}(h=K/\pi)$ are the kaon and pion yields in the kaon enriched sample with [$\mathcal{L}(K/\pi)>0.6$]. The pion misidentification rate $\kappa$ is a free parameter, as well as $R^{D}$ and $N^{D^{+}\pi^{-}}_{\rm total}$, where the latter is the total signal yield for the $\overline{B}{}^0\rightarrow D^{+}\pi^{-}$ decay. Due to a small contribution from $\overline{B}{}^0\rightarrow D^{+}K^{-}$ cross feed in the pion-enriched sample, the kaon identification efficiency $\epsilon_{K}$ is fixed to the value given in Sec.~\ref{sec:reco}. The yields are obtained from fitting the $\Delta E$ distribution. The background components are divided into the following categories in the fit:
\begin{enumerate}
    \item continuum $q\overline{q}$ background and combinatorial $B\overline{B}$ background, in which the final state particles could be from either the $B$ or $\overline{B}$ meson in an event; and 
    \item cross feed background from $\overline{B}{}^0\rightarrow D^{+}h^{-}$, where $h = \pi, K$, in which the charged kaon is misidentified as a pion or vice versa.
\end{enumerate} 
The $\overline{B}{}^0\rightarrow D^{+}h^{-} (h=K/\pi)$ signal distributions are represented by the sum of a double Gaussian function and  an asymmetric Gaussian with a common mean. These signal probability density functions (PDFs) are common to both kaon- and pion-enhanced samples. The means of the signal PDFs for $\overline{B}{}^0\rightarrow D^{+}\pi^{-}$ and $\overline{B}{}^0\rightarrow D^{+}K^{-}$  are directly extracted from the data, along with a single scaling factor to the narrowest signal Gaussian to account for any difference in $\Delta E$ resolution between simulated and data samples. Other parameters are fixed to those obtained from a fit to a large simulated sample of signal events.
 
A combined PDF is used to model combinatorial background consisting of continuum background and  $B\overline{B}$ background for  $\overline{B}{}^0\rightarrow D^{+}K^{-}$ ($\overline{B}{}^0\rightarrow D^{+}\pi^{-}$) decay, where the continuum is modeled with a first-order polynomial and the combinatorial $B\overline{B}$ background with an exponential function. The slope of the linear background and the exponential function’s exponent are determined from the fit to data; other parameters are fixed to those obtained from a fit to the corresponding simulated sample.
 
 The cross feed  background is described by a double Gaussian function in the $\overline{B}{}^0\rightarrow D^{+}K^{-}$ ($\overline{B}{}^0\rightarrow D^{+}\pi^{-}$) sample. The mean and scale factor for the $\overline{B}{}^0\rightarrow D^{+}\pi^{-}$ cross feed component PDF in the kaon-enhanced $\overline{B}{}^0\rightarrow D^{+}K^{-}$  sample are determined from the fit to data.
 
\begin{figure*}[htb!]
 \centering
\includegraphics[width=0.48\linewidth,page=1]{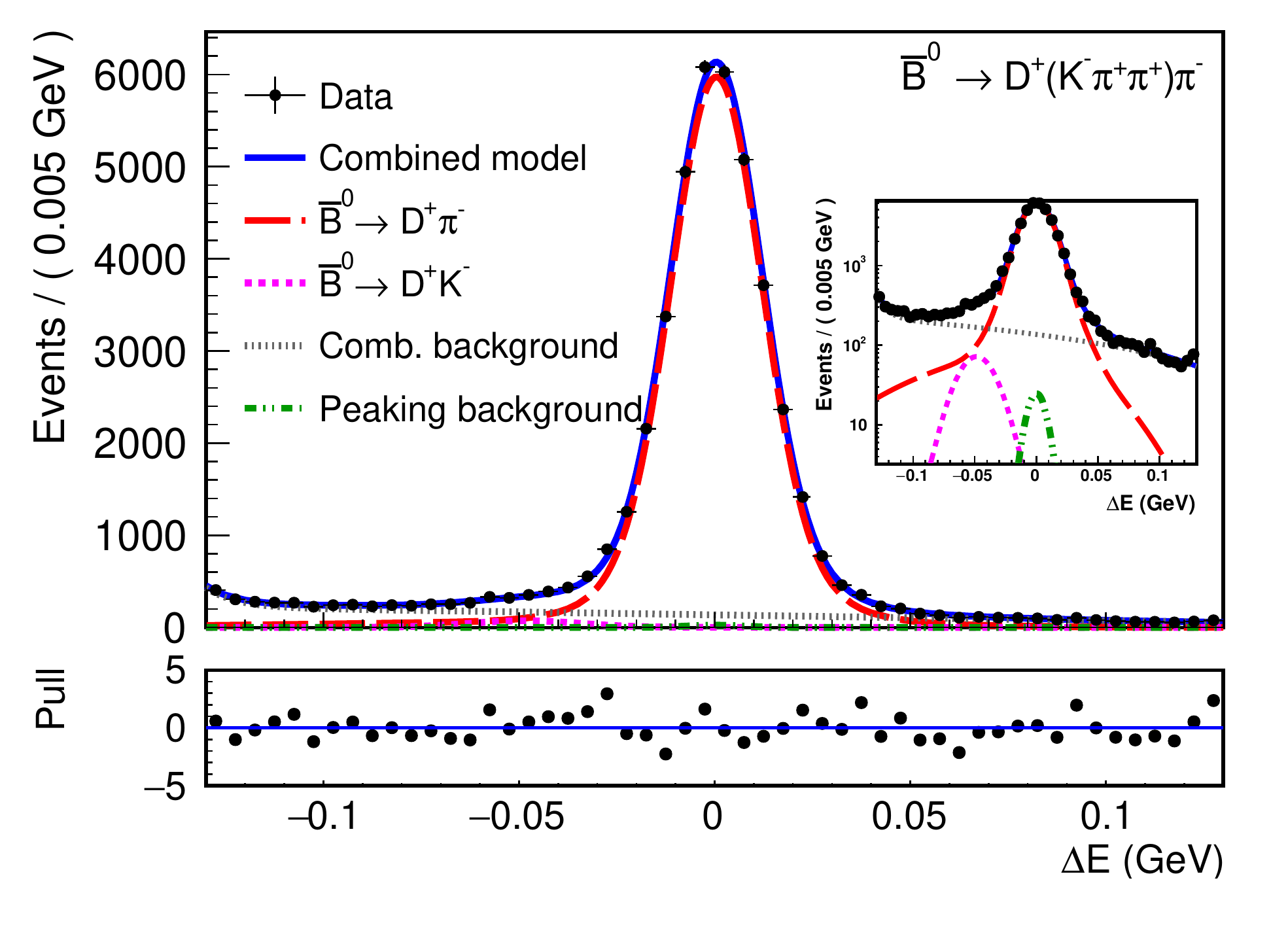} 
\includegraphics[width=0.48\linewidth,page=2]{data_nominal_fit_new.pdf} 
\caption{$\Delta E$ distributions for $\overline{B}{}^0\rightarrow D^{+}h^{-}$ candidates obtained from the (left) pion-enriched $\overline{B}{}^0\rightarrow D^{+}\pi^{-}$  and (right) kaon-enriched $\overline{B}{}^0\rightarrow D^{+}K^{-}$  data samples. The projections of the combined fit and individual components of a simultaneous unbinned maximum-likelihood fit are overlaid. The long-dashed red curve shows the $\overline{B}{}^0\rightarrow D^{+}\pi^{-}$ component.  The large-dotted magenta curve shows the $\overline{B}{}^0\rightarrow D^{+}K^{-}$  component. The small-dotted gray curve shows the combinatorial background component and the dash-dotted green curve show the peaking background component in $\overline{B}{}^0\rightarrow D^{+}\pi^{-}$  decay. The distribution of pulls between the fit and the data points is also shown.}
 \label{fig:Simultanenous_fit}
 \end{figure*}

There is a background  that can peak in the same manner as the $\overline{B}{}^0\rightarrow D^{+}\pi^{-}$ signal mode, which we call the ``peaking background''. The most prominent decay that peaks in the $\Delta E$ distribution is $B^{0} \rightarrow K^{*}J/\psi, K^{*} \rightarrow K^{+}\pi^{-}, J/\psi \rightarrow \mu^{+}\mu^{-}$  or $ e^{+}e^{-}$. This source accounts for $\sim$2$\%$ of the total background. To reject this contamination arising due to leptons misidentified as pions, we veto candidates with an invariant mass $M(\pi^+\pi^-)$ value falling within $\pm3\sigma$ of the known $J/\psi$ mass \cite{cite-PDG}. This essentially removes this peaking background  with $\sim$3$\%$ signal loss. The remaining peaking background contributions include semileptonic $D$ decays  for which the normalization is fixed from MC simulation. All yields are determined from a fit to data except for the peaking background yield. The uncertainty associated with the fixed peaking component is included in the systematic uncertainties. All other shape parameters are fixed to  their MC values. The yields obtained from the fit are listed in Table \ref{tab:results},  and the signal-enhanced fit projections for the data are shown in Fig.~\ref{fig:Simultanenous_fit}.
 \begin{table*}[!htb]
\caption{Various event yields and their statistical uncertainties obtained from the simultaneous fit.}
\centering
\renewcommand{\arraystretch}{1.6}
\begin{tabularx}{0.45\linewidth}{lc}
\hline \hline
~Parameter &~~~~~~~ Fit value ~~~~~~~\\
\hline
$\overline{B}{}^0\rightarrow D^{+}\pi^{-}$ total yield & $ 42065 \pm 235 $  \\
$\overline{B}{}^0\rightarrow D^{+}\pi^{-}$ background yield & $ 7414 \pm 128$   \\
$\overline{B}{}^0\rightarrow D^{+}K^{-}$ background yield & $ 2458 \pm 89$   \\
\hline
\end{tabularx}
\label{tab:results}
\end{table*}

\section{RESULTS} 
\label{sec:results}
The branching fraction of $\overline{B}{}^0\rightarrow D^{+}\pi^{-}$ decay is calculated as, 
\begin{eqnarray}
&&\mathcal{B}(\overline{B}{}^0\rightarrow D^{+}\pi^{-}) =\nonumber\\  
&&\frac{N^{\rm total}_{D^{+}\pi^{-}}}{2\times f_{00} \times N_{B\overline{B}} \times \epsilon_{D^{+}\pi^{-}} \times \mathcal{B}(D^{+}\rightarrow K^{-}\pi^{+}\pi^{+})},
\end{eqnarray}
where $N^{\rm total}_{D^{+}\pi^{-}}$ is the yield of $\overline{B}{}^0\rightarrow D^{+}\pi^{-}$ obtained  from the fit,  $N_{B\overline{B}}$ is the total number of $B\overline{B}$ pairs, $\epsilon_{D^{+}\pi^{-}} = (24.09 \pm 0.04)\%$ is the detection efficiency for $\overline{B}{}^0\rightarrow D^{+}\pi^{-}$ determined from signal MC events where the error is the associated statistical error from MC sample. The factor $f_{00}$ represents the neutral $B$ meson production ratio at the $\Upsilon (4S)$, which is $0.486\pm 0.006$ \cite{cite-PDG}, and  $\mathcal{B}(D^{+}\rightarrow K^{-}\pi^{+}\pi^{+})$ is the subdecay branching fraction of $D^{+}$, which is $(9.38\pm 0.16)\%$ \cite{cite-PDG}. The branching fraction for $\overline{B}{}^0\rightarrow D^{+}K^{-}$ decay is calculated by multiplying the $R^{D}$ value from the fit by the calculated $\overline{B}{}^0\rightarrow D^{+}\pi^{-}$ branching fraction.

The systematic uncertainties in the measurements from various sources are listed in Table~\ref{tab:R_syst}. Since the kinematics of $\overline{B}{}^0\rightarrow D^{+}\pi^{-}$ and  $\overline{B}{}^0\rightarrow D^{+}K^{-}$ processes are similar, most of the systematic effects cancel in the ratio of their branching fractions. The main source of systematic uncertainty that does not cancel is the uncertainty in $K/\pi$ identification efficiency. All the sources of systematic uncertainty are assumed to be independent, such that the total uncertainty is the quadratic sum of their contributions. The uncertainty associated with the $D^{+} \rightarrow K^{-}\pi^{+}\pi^{+}$ subdecay branching fraction is taken from its world average \cite{cite-PDG}. The uncertainty due to prompt tracking efficiency is based on a previous study of high momentum $(p> 200~\rm MeV)$ tracks. Tracking efficiency is calculated as the ratio between partially and fully reconstructed $D^{+}$ decays in data and MC events. The entry for  $N_{B\overline{B}}$ represents the uncertainty in the total number of $B\overline{B}$ events in data. Here $f_{00}$ refers to the uncertainty due to  $\mathcal{B} (\Upsilon(4S)\rightarrow B^{0}\overline{B}{}^{0})$ branching fraction calculated from PDG 2020 \cite{cite-PDG} along with the uncertainty due to isospin asymmetry calculated in \cite{cite-isospin}. The efficiency variation due to the $D^{+} \rightarrow K^{-}\pi^{+}\pi^{+}$ model is evaluated by varying the model and adding a phase space component. The resulting difference with respect to the measured central value of the branching fraction is treated as a systematic uncertainty. The systematic uncertainty due to  PDFs for the $D^{+}h^{-}(h=K/\pi)$ components and  the $D^{+}h^{-}(h=K/\pi)$ cross feed components are evaluated by varying the fixed shape parameters by $\pm 1\sigma$. The uncertainty due to the kaon identification efficiency is calculated by varying the measured value by its uncertainty obtained in data from the $D^{*}$ calibration sample as described in Sec.~\ref{sec:reco}.  The $D$ mass window and $M(\pi^{+}\pi^{-})$  for veto position have been varied and the resulting difference with respect to the measured branching fraction is taken as a systematic. The uncertainty due to the peaking background is obtained by varying its yield by the statistical uncertainty in its estimation.  The uncertainty associated with the reconstruction efficiency is measured using signal MC data samples.  We perform tests to validate the fit procedure and determine any possible bias in the fit procedure. The bias is not corrected and is used as a systematic uncertainty. The uncertainty due to the continuum suppression requirement is found to be negligible.
 
\begin{table*}[!htb]
\caption{Systematic uncertainties in the measured $R^{D}$ value and branching fractions for $\overline{B}{}^0\rightarrow D^{+}\pi^{-}$ and $\overline{B}{}^0\rightarrow D^{+}K^{-}$. The total systematic uncertainty is the quadratic sum of the uncorrelated uncertainties.}
\centering
\renewcommand{\arraystretch}{1.6}
\begin{tabularx}{0.66\linewidth}{lccc}
\hline \hline
Source~~~~ & ~~~~$R^{D}$  ~~~~& ~~~~$\mathcal{B}(\overline{B}{}^0\rightarrow D^{+}\pi^{-})$ ~~~~& ~~~~ $\mathcal{B}(\overline{B}{}^0\rightarrow D^{+}K^{-})$\\ \hline
$\mathcal{B}(D^{+}\rightarrow K^{-}\pi^{+}\pi^{+})$ 		&-- & 1.71$\%$ &1.71$\%$\\ \hline
Tracking 									 & --& 1.40$\%$  & 1.40$\%$\\
$N_{B\overline{B}}$ 							& -- &  1.37$\%$&1.37$\%$ \\ 
$f^{00}/f^{+-}$  				& -- & 1.92$\%$ & 1.92$\%$ \\ 
$D^{+}\rightarrow K^{-}\pi^{+}\pi^{+}$ model & -- & $0.69\%$  & $0.69\%$\\
PDF parametrization  		& $2.71\%$ & 1.63$\%$  & $1.79\%$\\
PID efficiency of  $K/\pi$  	& $0.88\%$ & $0.68\%$   & $0.73\%$ \\  
$D^{+}$ mass selection window  &$0.05\%$ & 0.56$\%$ & $0.64\%$\\
$J/\psi$ veto selection  &$0.12\%$ & 0.004$\%$  & $0.15\%$\\
Peaking background yield &  $0.07\%$ & 0.04$\%$  & $0.00\%$\\
MC statistics  		& $<0.01$& 0.04$\%$ & 0.04$\%$\\
Fit bias		&-- &0.58$\%$ & $0.61\%$\\
\hline
Total 	&$2.85\% $ & $3.43\%$  &$3.54\%$\\ \hline
\end{tabularx}
\label{tab:R_syst}
\end{table*}
The ratio of branching fractions is found to be,
\begin{eqnarray}
\label{eq:r}
R^{D}    &=& 0.0819 \pm 0.0020 (\rm stat) \pm 0.0023(\rm syst).
\end{eqnarray}
The total $D^{+}\pi^{-}$ yield from the simultaneous fit is used to determine the branching fraction of the $\overline{B}{}^0\rightarrow D^{+}\pi^{-}$ decay, 
\begin{align}
\mathcal{B}(\overline{B}{}^0\rightarrow D^{+}\pi^{-}) & = (2.48 \pm 0.01 \pm 0.09  \pm 0.04) \times 10^{-3} \notag \\
  & \label{eq:dpi}
\end{align}
 where the first uncertainty is statistical, the second is systematic, and the third is associated with $D^{+} \rightarrow K^{-}\pi^{+}\pi^{+}$ branching fraction. The branching fraction of $\overline{B}{}^0\rightarrow D^{+}K^{-}$ is calculated by multiplying Eq. \eqref{eq:r}  by Eq. \eqref{eq:dpi},
\begin{multline}
\label{eq:dk}
\mathcal{B}(\overline{B}{}^0\rightarrow D^{+}K^{-})= (2.03 \pm 0.05 \pm 0.07 \pm 0.03) \times 10^{-4}
\end{multline}
The $\kappa$ value obtained from the fit is $(7.79 \pm 0.21)\%$, which agrees within one standard deviations with the expected pion misidentification rate as given in Sec.~\ref{sec:reco}. In both measurements listed in Eqs. \eqref{eq:dpi} and \eqref{eq:dk}, one of the dominant sources of systematic uncertainty arises from the fixed PDF parametrization. 
 \section{Conclusion} 
 \label{sec:conclusion}
In summary, we have reported measurements of the branching fraction ratio between Cabibbo suppressed $\overline{B}{}^0\rightarrow D^{+}K^{-}$ and Cabibbo favored $\overline{B}{}^0\rightarrow D^{+}\pi^{-}$ using the full $\Upsilon(4S)$ data sample collected by the Belle experiment, which supersedes the previous Belle measurement \cite{cite-belleold}. We also present a measurement of the branching fractions for $\overline{B}{}^0\rightarrow D^{+}\pi^{-}$ and $\overline{B}{}^0\rightarrow D^{+}K^{-}$ decays.The $\overline{B}{}^0\rightarrow D^{+}h^{-}(h=K/\pi)$ branching fraction and $R^{D}$ values are compatible with the corresponding world averages \cite{cite-PDG} within their uncertainties. Individual branching fractions of $\overline{B}{}^0\rightarrow D^{+}\pi^{-}$ and $\overline{B}{}^0\rightarrow D^{+}K^{-}$ deviate from the theory predictions in Refs. \cite{cite-arXix1606.02888, cite-arXiv:2007.10338}, however, the ratio agrees within uncertainties.

\section*{Acknowledgements}
We thank the KEKB group for the excellent operation of the
accelerator; the KEK cryogenics group for the efficient
operation of the solenoid; and the KEK computer group, and the Pacific Northwest National
Laboratory (PNNL) Environmental Molecular Sciences Laboratory (EMSL)
computing group for strong computing support; and the National
Institute of Informatics, and Science Information NETwork 5 (SINET5) for
valuable network support.  We acknowledge support from
the Ministry of Education, Culture, Sports, Science, and
Technology (MEXT) of Japan, the Japan Society for the 
Promotion of Science (JSPS), and the Tau-Lepton Physics 
Research Center of Nagoya University; 
the Australian Research Council including Grants No.
DP180102629, 
No. DP170102389, 
No. DP170102204, 
No. DP150103061, 
and No. FT130100303; 
Austrian Federal Ministry of Education, Science and Research (FWF) and
FWF Austrian Science Fund No.~P~31361-N36;
the National Natural Science Foundation of China under Contracts
No.~11435013,  
No.~11475187,  
No.~11521505,  
No.~11575017,  
No.~11675166,  
and No.~11705209;  
Key Research Program of Frontier Sciences, Chinese Academy of Sciences (CAS), Grant No.~QYZDJ-SSW-SLH011; 
the  CAS Center for Excellence in Particle Physics (CCEPP); 
the Shanghai Science and Technology Committee (STCSM) under Grant No.~19ZR1403000; 
the Ministry of Education, Youth and Sports of the Czech
Republic under Contract No.~LTT17020;
Horizon 2020 ERC Advanced Grant No.~884719 and ERC Starting Grant No.~947006 ``InterLeptons'' (European Union);
the Carl Zeiss Foundation, the Deutsche Forschungsgemeinschaft, the
Excellence Cluster Universe, and the VolkswagenStiftung;
the Department of Atomic Energy (Project Identification No. RTI 4002) and the Department of Science and Technology of India; 
the Istituto Nazionale di Fisica Nucleare of Italy; 
National Research Foundation (NRF) of Korea Grants 
No.~2016R1\-D1A1B\-01010135, No. 2016R1\-D1A1B\-02012900, No. 2018R1\-A2B\-3003643,
No. 2018R1\-A6A1A\-06024970, No. 2019K1\-A3A7A\-09033840,
No. 2019R1\-I1A3A\-01058933, No. 2021R1\-A6A1A\-03043957,
No. 2021R1\-F1A\-1060423, No. 2021R1\-F1A\-1064008;
Radiation Science Research Institute, Foreign Large-size Research Facility Application Supporting project, the Global Science Experimental Data Hub Center of the Korea Institute of Science and Technology Information and KREONET/GLORIAD;
the Polish Ministry of Science and Higher Education and 
the National Science Center;
the Ministry of Science and Higher Education of the Russian Federation, Agreement 14.W03.31.0026, 
and the HSE University Basic Research Program, Moscow; 
University of Tabuk research Grants No.
S-1440-0321, No. S-0256-1438, and No. S-0280-1439 (Saudi Arabia);
the Slovenian Research Agency Grants No. J1-9124 and No. P1-0135;
Ikerbasque, Basque Foundation for Science, Spain;
the Swiss National Science Foundation; 
the Ministry of Education and the Ministry of Science and Technology of Taiwan;
and the United States Department of Energy and the National Science Foundation.

\end{document}